\documentclass{aastex}          
\usepackage{spr-astr-addons}    
\usepackage{url}
\newcommand{\emaila}{goicol@unican.es}

\begin{document}
%
\title{Crowded-field image simulator for WSO-UV/ISSIS: first functional version developed by 
the Glendama team}

\shorttitle{Crowded-field Image Simulator}
\shortauthors{Shalyapin \& Goicoechea}

\author{V. N. Shalyapin\altaffilmark{1}} 
\and 
\author{L. J. Goicoechea}
\affil{Departamento de F\'\i sica Moderna, Universidad de Cantabria, 
Avda. de Los Castros s/n, 39005 Santander, Spain}
\email{\emaila} 

\altaffiltext{1}{Institute for Radiophysics and Electronics, National Academy 
of Sciences of Ukraine, 12 Proskura St., 61085 Kharkov, Ukraine}

\begin{abstract}
We are developing a web-based interactive software to simulate crowded-field imaging with ISSIS 
on board the future WSO-UV. This new tool is aimed to prepare WSO-UV/ISSIS proposals to observe 
multicomponent targets and dense fields. For a given combination of UV channel, filters and 
exposure time, the user creates a set of point-like and extended sources (source model). This 
source model produces a final image, which takes into account a pixelated field of view, a 
realistic conversion between physical flux and counts per second, the convolution with the 
expected point spread function, a sky background and noise fluctuations. The current version of 
the simulator is available at the Glendama website, and it allows users to specify all relevant 
parameters of each point-like or extended source, drag-and-drop sources by using a mouse or a
fingertip/stylus on a touchscreen, change the frame size or the brightness scale, etc. 
\end{abstract}


\section{Introduction}
The Imaging and Slitless Spectroscopy Instrument for Surveys \citep[ISSIS;][]{gomez12,gomez14} 
will be one of the two main science units (the WUVS spectrographs are the other) on board the 
World Space Observatory-Ultraviolet (WSO-UV) mission \citep{sachkov14a,sachkov14b,shustov14}, 
which is intended to be launched in 2017. This UV space observatory will incorporate an 1.7 m 
telescope, and it is expected to operate for about 10 years. 

The imaging mode of ISSIS should allow astronomers to resolve the fine structure of faint UV 
sources, as well as multiple faint sources in dense fields. This is because the small pixel 
scale, compact point spread function (PSF) and high sensitivity of the instrument 
\citep{gomez14}. Each ISSIS image ($2048 \times 2048$ pixels) or sub-image will be the 
convolution of a brightness distribution (generated by sources inside the corresponding field 
of view) with the PSF. The sky background and noise fluctuations will also play a role. 

\begin{figure}[t]
  \includegraphics[width=\columnwidth]{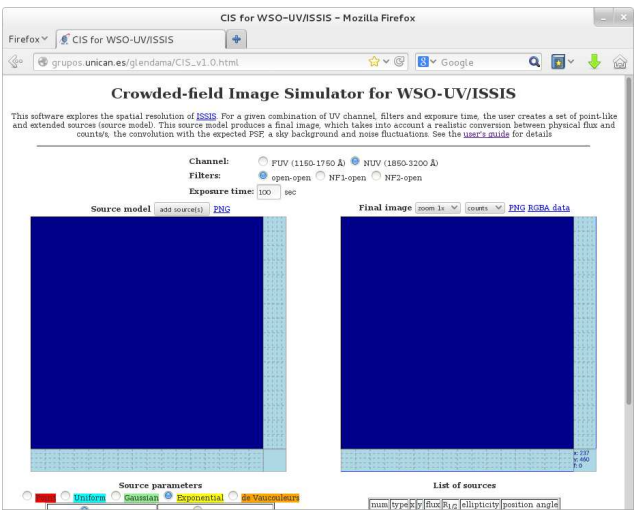}
\caption{Top part of the simulator main window. The bottom part contains controls 
to set sources, as well as the list of sources.}
\label{main_window}
\end{figure}

To accurately model ISSIS sub-images of crowded fields, we are making a free, web-based 
interactive software (see Fig.~\ref{main_window}). In Sect. 2 and Sect. 3 we give detailed 
instructions for potential users of our crowded-field image simulator (CIS), while in Sect. 4 
we present some examples of using the CIS. Sect. 5 includes some final remarks on the new tool 
and its applicability to study gravitational lens systems in the UV.

\section{User Guide}
The CIS is available at the Glendama website. Click on the URL: 
\url{http://grupos.unican.es/glendama/CIS.html} to start this simulator. Your browser will 
open a main window that contains empty frames of $512 \times 512$ (or $256 \times 256$) pixels 
covering $18.4\arcsec \times 18.4\arcsec$ (or $9.2\arcsec \times 9.2\arcsec$) on the sky (see 
Fig.~\ref{main_window}). In the current version, the initial frame dimensions depend on the size 
of your screen, and the pixel scale is 0.036\arcsec. Later you can reduce the field of view and 
go into detail by using the zoom button. In Fig.~\ref{main_window}, the bottom horizontal and 
right vertical panels in light blue trace fluxes along the row and column for the position of 
the pointer, respectively.

The first step is to select a channel (FUV or NUV). Do not forget its spectral range (1150$-$1750 
\AA\ and 1850$-$3200 \AA\ for the FUV and NUV channels) when entering fluxes, since we consider 
smooth spectrum sources that are approximated by flat-spectrum objects with averaged fluxes. This 
allows users to avoid spectral details, and focus on spatial and structural features. You should 
also set the filter wheel configuration (open = no filter, NF1 = neutral filter with a constant 
transmittance of about 0.1, NF2 = neutral filter with a constant transmittance of about 0.01) and 
the exposure time.
 
In order to add sources in the left-hand frame, you must first choose a source type. Point-like 
sources are displayed as red stars in the left-hand frame, while uniform, Gaussian, exponential 
or de Vaucouleurs (extended) sources are shown as cyan, light green, yellow and orange ellipses, 
respectively. In a next step, you have two options: {\it specific object} or {\it random 
distribution}.
 
 \begin{enumerate}
 \item 
 {\it Specific object}: You can add one source, indicating its position ($x$, $y$; both in pixels), 
 flux ($F$ in erg cm$^{-2}$ s$^{-1}$ \AA$^{-1}$) and structure parameters. For a point-like object, 
 you can type/paste any value of these last parameters, since they are ignored by the simulator. 
 However, an extended source with elliptical isophotes is characterised by the major radius of 
 the ellipse enclosing 50\% of the total light ($R_{1/2}$ in pixels), as well as its ellipticity 
 ($\epsilon = 1 - b/a$, where $a$ and $b$ are the semi-major and semi-minor axes) and 
 orientation ($PA$ in degrees) of isophotes. Once you enter all data, click on the "add source(s)" 
 button.        
 \item 
 {\it Random distribution}: Enter the number of sources (1, 10 or 100), and set the lower and 
 upper limits of fluxes (in erg cm$^{-2}$ s$^{-1}$ \AA$^{-1}$) for point-like objects. For 
 extended objects, you also need to assign the ranges to their structure parameters $R_{1/2}$,
 $\epsilon$ and $PA$ (see above). Click on the "add source(s)" button.
 \end{enumerate}

You can put the pointer on any point-like or extended source to see its parameters: $x$, $y$, 
$F$ (point-like) or $x$, $y$, $F$, $R_{1/2}$, $\epsilon$, $PA$ (extended). You can add as 
many sources as necessary, change the position of any object (click on it, drag and drop) and 
remove it (double click).
 
The source model leads to a final image in the right-hand frame. First, physical fluxes (in erg 
cm$^{-2}$ s$^{-1}$ \AA$^{-1}$) are converted to counts per second (cps) using a realistic 
cps-to-flux ratio for the corresponding channel, i.e., 3.36 $\times$ 10$^{15}$ (FUV) or 1.14 
$\times$ 10$^{17}$ (NUV). These ratios are properly reduced when using a neutral filter (see 
above for transmittance factors). Second, the source model is convolved with the expected central 
PSF for the FUV/NUV channel. Third, a sky background is added to the image. This is due to the 
Earthshine (ES), the Zodiacal light (ZL) and the Geocoronal emission (GC), and we take the ES, ZL 
and GC average fluxes in the Hubble Space Telescope (HST)\footnote{See documents on the HST UV 
background at \url{http://www.stsci.edu/hst}}. 

The ES flux for the WSO-UV telescope should have smaller values than those for the HST, because the 
higher Earth orbit of the WSO-UV (less affected by the reflected sunlight). However, although the 
ES contribution is overestimated, the total background (based on the ES, ZL and GC average fluxes 
in the HST) is dominated by the GC signal: ES $<$ ZL $<$ GC. This sky background depends on the 
selected channel and filters, and its open-open values are $\beta = 4.9 \times 10^{-3}$ cps/pix (ES 
+ ZL = $7.2 \times 10^{-12}$ cps/pix) in the FUV and $\beta = 3.5 \times 10^{-4}$ cps/pix (ES + ZL 
= $2.1 \times 10^{-5}$ cps/pix) in the NUV. The current version of the simulator does not account 
for any instrumental background (e.g., dark current).
 
In a final step, the simulator takes into account the exposure time ($T$ in sec) to produce a 
number of counts in each pixel, and then generate Poisson fluctuations. The final image initially
displays counts in pixels ($C$), but the user is allowed to modify the brightness scale, i.e., 
$\log(C)$ or $\sqrt{C}$ instead of $C$, or zoom the image. We note that the logarithmic scale is 
related to magnitudes. In addition, the signal-to-noise ratio (per pixel) is given by SNR = $(C - 
B)/\sqrt{C}$, where $B = \beta \times T$. Hence, the square-root scale is useful to quantify the 
SNR when $C >> B$. 
 
\section{Saving and handling simulations}
You can save the source model and the final image to your computer by clicking on the "PNG" labels 
above both frames. After each click, your browser will open an additional window displaying the 
corresponding PNG file. This file can be stored in a suitable folder using the browser command 
"File/Save as...". 

Regarding the final image, it is also possible to display on your screen and save a 
red-green-blue-alpha (RGBA) data file. This RGBA data file contains information to reconstruct the 
simulated brightness map. The $C$ values are described from the 3 colour (RGB) indexes. The opacity 
(A) index does not play a role, and 
\begin{equation}
 C = \rm{R} + 256 \times \rm{G} + 256 \times 256 \times \rm{B} .
\label{Eq1} 
\end{equation}
Here below we describe a simple way to reconstruct the final image in standard FITS format. This
recipe only relies on commonly-used astronomical software (DS9 and IRAF). 

{\it Recipe for cooking data}: Click on the "RGBA data" label to show this file on your screen. 
You can then save it in PNG format, e.g., image.png. Open your DS9 viewer and select "Frame/New 
Frame RGB" and "File/Import/PNG..." Then you should import the file of interest (image.png). In 
the RGB panel of DS9, select "Current/Red" and save the corresponding FITS file, i.e., 
"File/Save/R.fits". Then select "Current/Green" and save G.fits, and also select "Current/Blue" and 
save a third file called B.fits. Now you need to combine the three FITS files to obtain the 
simulated brightness map as a FITS image. This task can be done in an IRAF environment. Taking the 
Eq. (\ref{Eq1}) into account, the IRAF commands are:\\
\\
cl\textgreater imar G * 256 G \\
cl\textgreater imar B * 65536 B \\
cl\textgreater imar R + G RG \\
cl\textgreater imar RG + B image .\\
\\
The final product (image.fits) is ready to be analysed in detail. You may use your favourite 
analysis software to explore the scientific return of the simulated observations, and thus, check 
the ISSIS performance and prepare observation proposals.

\begin{figure}[t]
\centering
  \begin{tabular}{cc}
    \includegraphics[width=0.45\columnwidth]{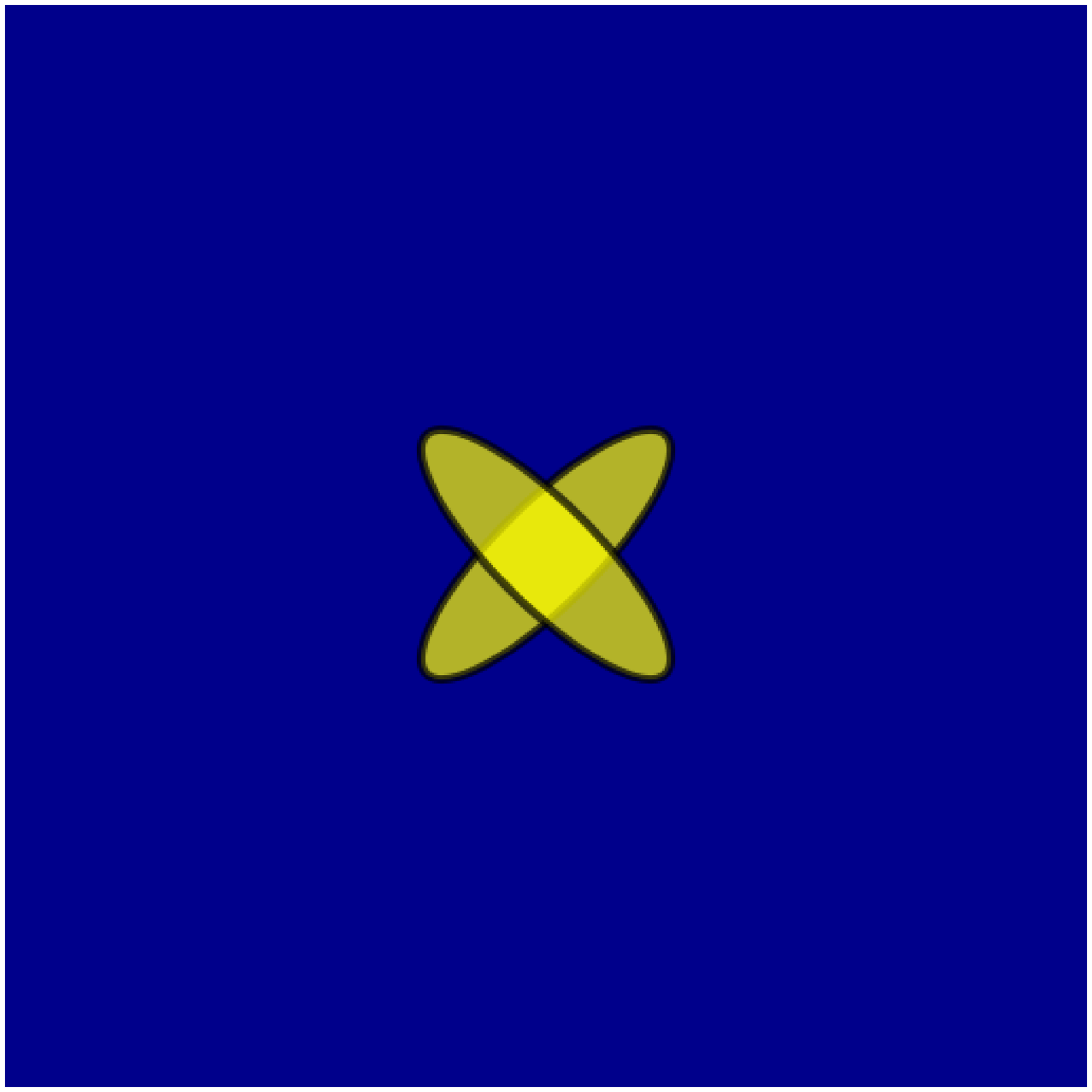} &
    \includegraphics[width=0.45\columnwidth]{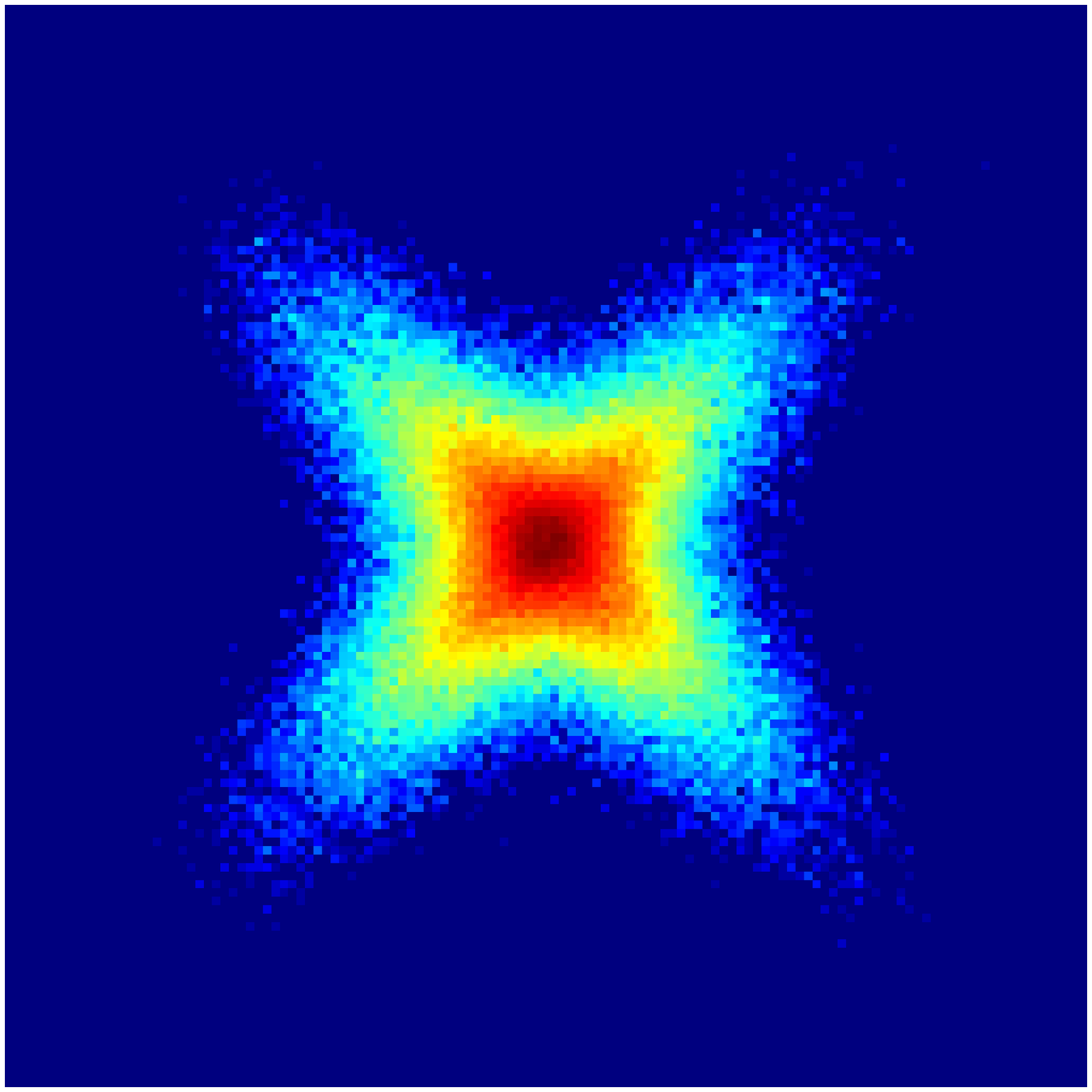}\\
  \end{tabular}
\caption{ISSIS sub-image of $128 \times 128$ pixels. {\it Left}: Two galaxies with exponential 
profiles. Both have the same central position ($x = y = 256$), flux (10$^{-14}$ erg cm$^{-2}$ 
s$^{-1}$ \AA$^{-1}$), $R_{1/2}$ value (20 pix, or equivalently, 0.72\arcsec) and ellipticity (0.7). 
However, their position angles differ by 90 deg. {\it Right}: Final image using the default 
configuration: NUV channel, open-open, $T$ = 100 sec. We also use the logarithmic brightness 
scale.}
\label{exp2}
\end{figure}

\begin{figure}[t]
\centering
  \begin{tabular}{cc}
    \includegraphics[width=0.45\columnwidth]{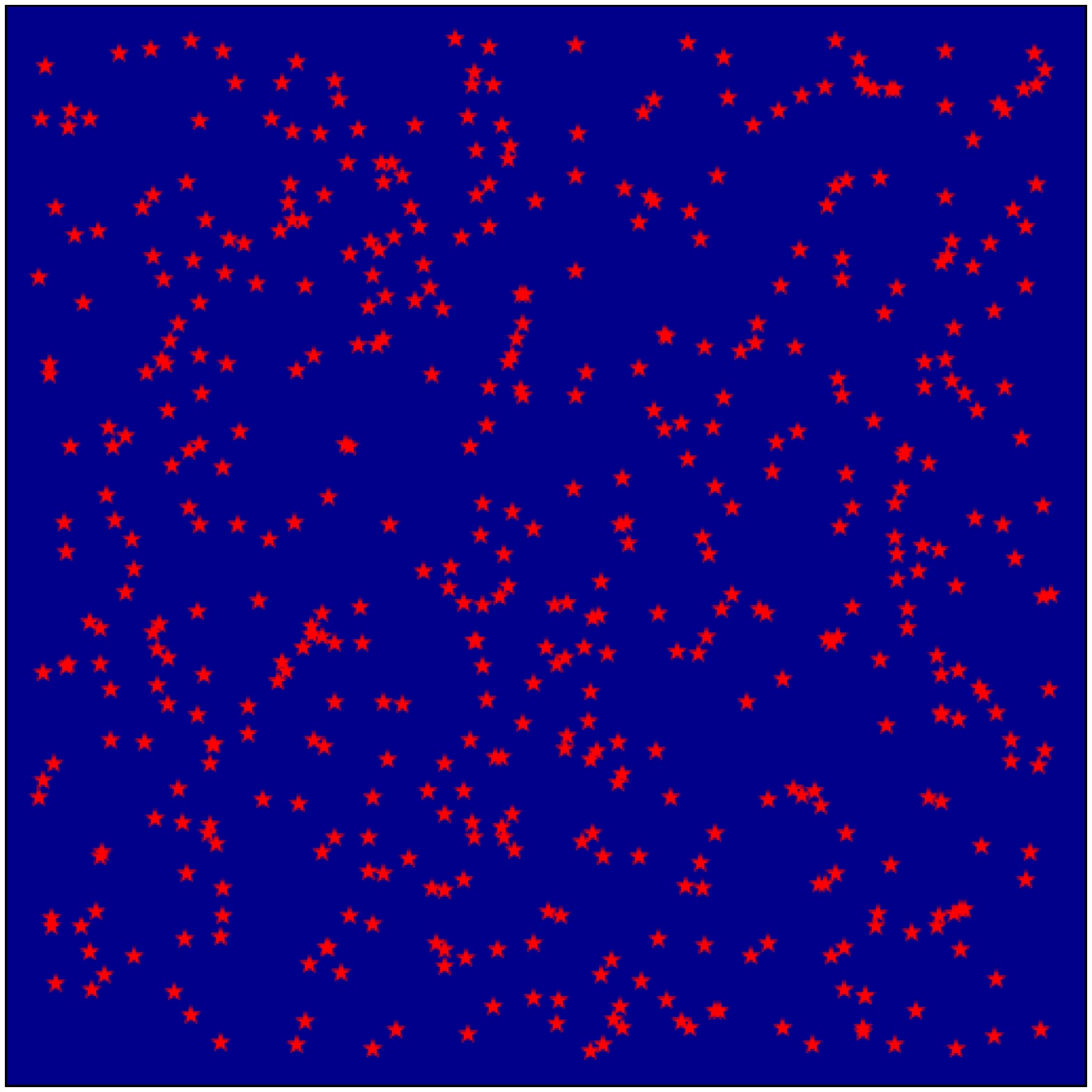} &
    \includegraphics[width=0.45\columnwidth]{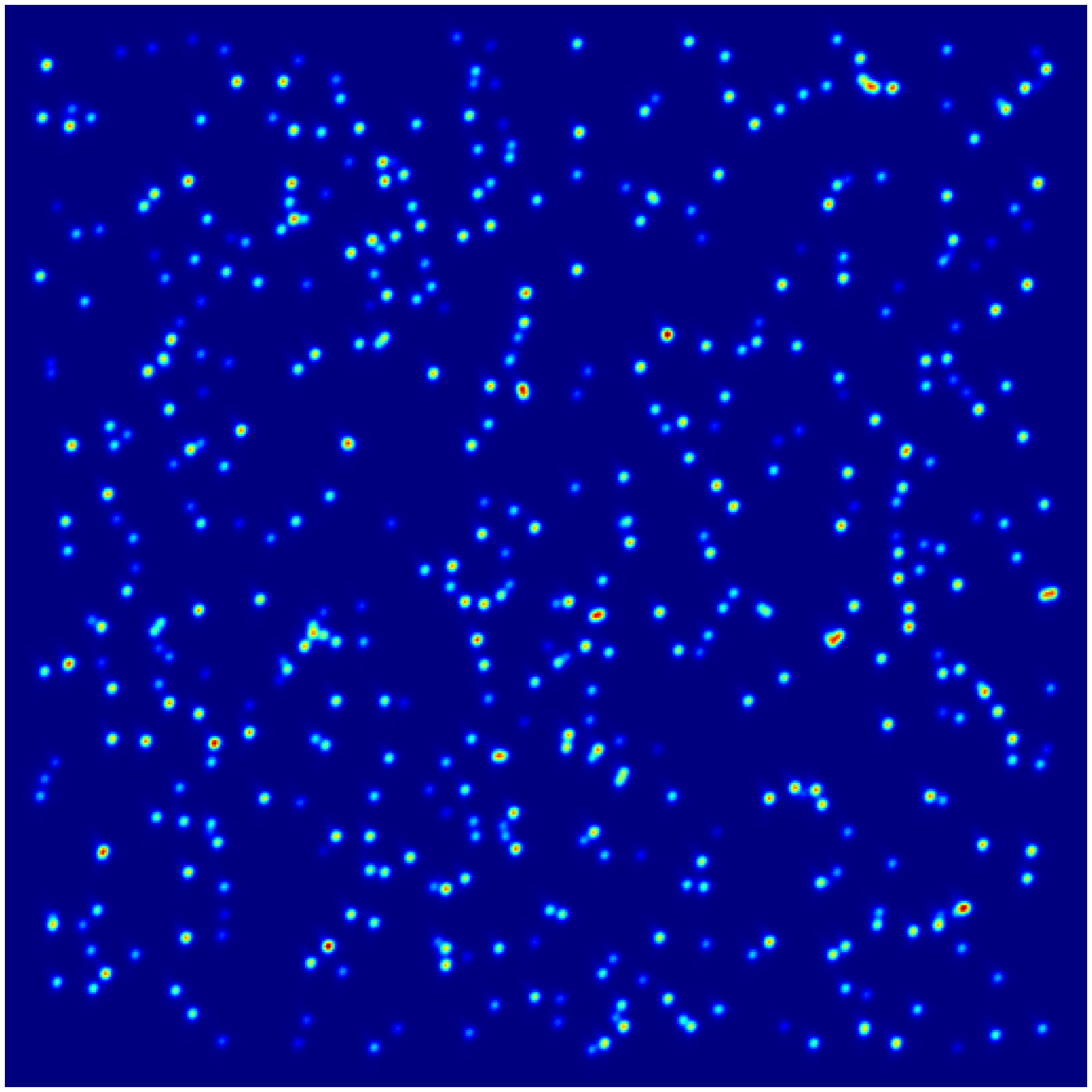}\\
  \end{tabular}
\caption{ISSIS sub-image of $512 \times 512$ pixels. {\it Left}: 500 random stars. These stars 
have fluxes in the default interval $10^{-15}-10^{-14}$ erg cm$^{-2}$ s$^{-1}$ \AA$^{-1}$. {\it 
Right}: Final image (counts) using the default configuration of the instrument (see the caption of 
Fig.~\ref{exp2}).}
\label{star500}
\end{figure}
 
\begin{figure}[t]
\centering
  \begin{tabular}{cc}
    \includegraphics[width=0.45\columnwidth]{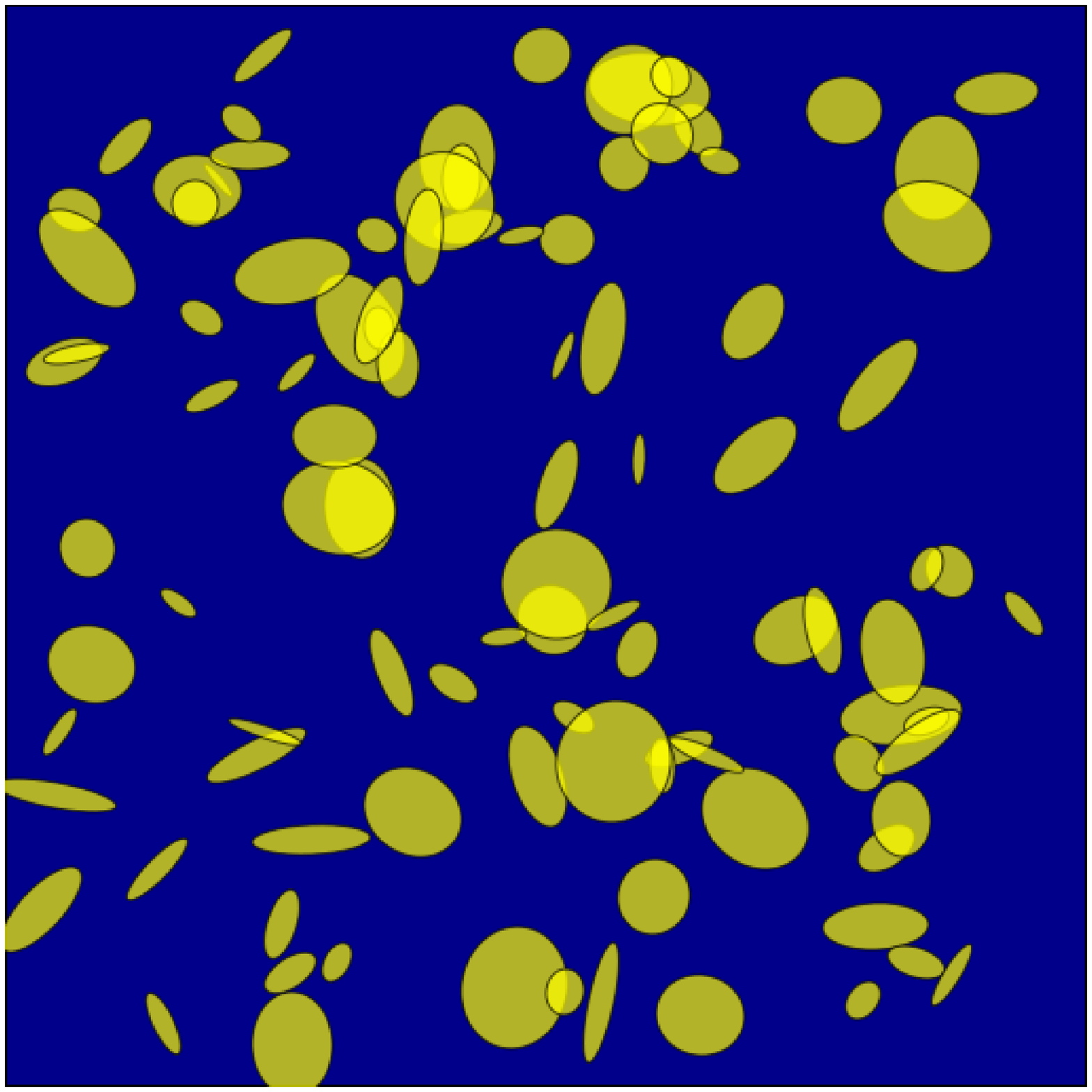} &
    \includegraphics[width=0.45\columnwidth]{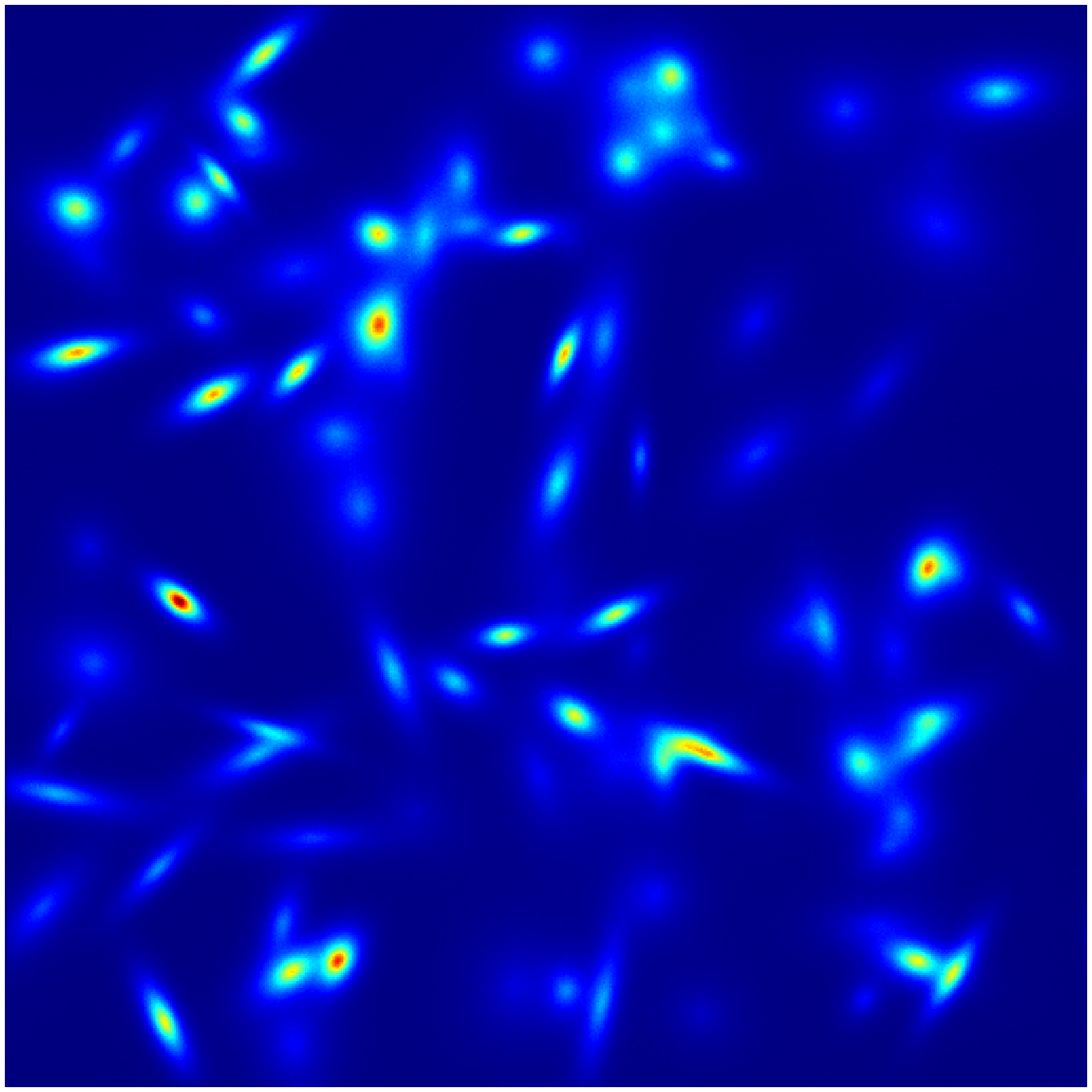}\\
  \end{tabular}
\caption{ISSIS sub-image of $512 \times 512$ pixels. {\it Left}: 100 random galaxies with 
exponential brightness profiles. Fluxes and structure parameters are within the default intervals: 
10$^{-15} \leq F$ (erg cm$^{-2}$ s$^{-1}$ \AA$^{-1}$) $\leq 10^{-14}$, 10 $\leq R_{1/2}$ (pixels) 
$\leq$ 30, 0 $\leq \epsilon \leq$ 0.8 and 0 $\leq PA$ (deg) $\leq$ 180. {\it Right}: Final image 
(counts) using the default configuration of the instrument (see the caption of Fig.~\ref{exp2}).} 
\label{exp100}
\end{figure}

\section{Simulation examples}
This section describes the steps to simulate a few crowded fields. In each simulation, we start with 
empty frames of $512 \times 512$ pixels (see Sect. 2). 

\begin{enumerate}
\item {\it Two galaxies that are blended together} 
  \begin{itemize}
    \item go to the "Source parameters/specific" table and put the cursor in the "ellipticity" box. 
    Enter an $\epsilon$ value of 0.7 (its default value is 0.5); 
    \item click the "add source(s)" button;
    \item change the "position angle (deg)" from 45 to 135;
    \item click the "add source(s)" button again;
    \item choose "zoom 4x" ($128 \times 128$ pixels) and "log (C)" in the drop-down menus of the 
    final image.
  \end{itemize}
Both the source model and the final image are shown in Fig.~\ref{exp2}. As an additional experiment, 
you may change the exposure time from its default value (100 sec) to shorter and longer values, e.g., 
10 and 1000 sec. This shorter/longer exposure would produce a clear worsening/improvement of the 
SNR in the final image [Hint: use "sqrt (C)" in the drop-down menu for the brightness scale].  
\item {\it 500 random stars} 
  \begin{itemize} 
    \item select "Source parameters/Point" and the "random" option;
    \item choose the value 100 in the drop-down menu for the "number of sources";
    \item click 5 times on the "add source(s)" button.
   \end{itemize}
In order to explore the star shape in the final image (see the right panel in Fig.~\ref{star500}), 
you can use the zoom control.  
\item {\it 100 random galaxies}  
  \begin{itemize} 
    \item select the "Source parameters/random" option;
    \item choose the value 100 in the drop-down menu for the "number of sources";
    \item click on the "add source(s)" button.
   \end{itemize}
You can find details on the random sample of galaxies in the "List of sources" (see also 
Fig.~\ref{exp100}). 
\end{enumerate} 

\section{Concluding remarks}
The World Space Observatory-Ultraviolet is a cosmic mission that will provide data about a large 
variety of targets. Many of these targets would have complex structures and/or would be located in 
dense fields, and thus, the crowded-field image simulator for ISSIS is expected 
to play a relevant role in the development of observation strategies. This new publicly available 
tool is designed to be powerful, yet user-friendly and intuitive. Users do not need to install the 
simulator, and only a web-browser is required to use the tool. If the software at the Glendama 
website is downloaded to a local computer, then the simulations can be also performed in an off-line 
mode (without Internet connection). 

Thanks to the implementations in JavaScript language, the new simulator combines simplicity and 
functionality. In recent years this language is evolving significantly faster than in its origins, 
and is being converted into a powerful calculation tool. For example, we use a 2-dimensional fast 
Fourier transform (2D FFT) on large grids, which was hard to imagine several years ago. At present,
other astronomy teams are actively developing JavaScript software or transferring desktop 
applications to web with JavaScript (e.g., the JS9 web implementation of the well-known DS9 FITS 
viewer). These tasks will be accelerated during next years and our simulator is one contribution to 
this ongoing process.         

The current version of the simulator does not incorporate all filters that will be available. Hence, 
we plan to take into account the different filter responses, and incorporate this information into 
a future version of the simulation tool. We also want to improve the adopted background and make 
easier (if possible) the obtention of final images in FITS format (to subsequent analyses). This 
final version should be available to the community before the first calls for proposals and the 
launch of the WSO-UV mission in 2017. 

We belong to the WSO-UV/ISSIS Science Working Group, and are involved in the preparation of a large
observation program to analyse optically-bright lensed quasars in the UV \citep{goico11}\footnote{See 
updates at \url{http://grupos.unican.es/glendama/}}. In more detail, using the 
CASTLES\footnote{\url{http://www.cfa.harvard.edu/castles/}} 
and 
SQLS\footnote{\url{http://www-utap.phys.s.u-tokyo.ac.jp/~sdss/sqls/}} databases, we selected 41 
systems in the redshift interval 1 $< z <$ 3. The key idea is to carry out two subprograms with this 
sample or part of it. The first subprogram aims to build an UV database (disentangle nuclear and 
circumnuclear EUV emissions, etc), while the second one focuses on an UV monitoring (variability of 
the nuclear continua). Although preliminary estimations support the feasibility of both programs, we 
will use the final simulator to check all details and apply for observing time on the future space 
observatory.

\acknowledgments
This research has been supported by the Spanish Department of Science and Innovation 
grant AYA2010-21741-C03-03 (Gravitational LENses and DArk MAtter - GLENDAMA project), 
and the University of Cantabria. 

%

\end{document}